\documentclass[twocolumn,showpacs,preprintnumbers,amsmath,aps,amssymb]{revtex4-1}
\usepackage{graphicx}
\usepackage{dcolumn}
\usepackage{bm}
\usepackage{color}

\begin{document}

\title{Effects of deformed phase space on scalar field cosmology.}
\author{S. P\'erez-Pay\'an}
\email{sinuhe@fisica.ugto.mx}
\author{M. Sabido}
\email{msabido@fisica.ugto.mx}
\affiliation{ Departamento  de F\'{\i}sica de la Universidad de Guanajuato,\\
 A.P. E-143, C.P. 37150, Le\'on, Guanajuato, M\'exico
 }%

\author{C. Yee-Romero}
 \email{carlos.yee@uabc.edu.mx}
\affiliation{  Departamento de Matem\'aticas, Facultad de Ciencias\\
 Universidad Aut\'onoma de Baja California, Ensenada, Baja California, M\'exico\\}%

\begin{abstract}
The effects of phase space deformations in standard scalar field cosmology  are studied. The deformation is introduced by modifying the symplectic structure of the minisuperspace variables to have a deformed Poisson algebra among the coordinates and the canonical momenta. It is found that in the deformed minisuperspace model  gives rise to an accelerating scale factor in the absence of a cosmological constant, and this acceleration is a consequence of the phase space deformation parameter $\beta$.
\end{abstract}
 \maketitle
 \newpage

\section{Introduction}
The initial interest in noncommutative field theory \cite{nekra} slowly but steadily  permeated in the realm of gravity, from which  several approaches to noncommutative gravity \cite{GarciaCompean:2003pm} where proposed. All of these formulations showed that the end result of a noncommutative theory of gravity, is a highly nonlinear theory. In order to study the effects of noncommutativity on different aspects of the universe, noncommutative cosmology was presented in  \cite {GarciaCompean:2001wy}. The authors noticed   that the noncommutative deformations modify the noncommutative fields, and  conjectured that the effects of the  full  noncummutative theory of gravity should be reflected in the minisuperspace variables.
 This was achieved  by introducing the Moyal product of functions in the Wheeler-DeWitt equation, in the same manner as is done in noncommutative quantum mechanics. The model analyzed  was the Kantowski-Sachs cosmology and was carried out at the quantum level, several works followed with this idea  \cite{ Barbosa:2004kp, Pimentel:2004jv}.

Although the noncommutative deformations of the minisuperspace where originally analyzed at the quantum level, by an effective noncommutativity on the minisuperspace, classical noncommutative formulations have been proposed. In  \cite{ Barbosa:2004kp}, the authors  considered  classical noncommutative relations in the phase space for the Kantowski-Sachs cosmological model and  established the classical noncommutative equations of motion. For scalar field cosmology, in \cite{Pimentel:2004jv, Guzman:2007} the  classical minisuperspace  is deformed and a scalar field is used as the matter component of the universe.
The main idea of this classical noncommutativity is based on the assumption that modifying the Poisson brackets of the classical theory gives the noncommutative equations of motion \cite{GarciaCompean:2001wy,Pimentel:2004jv,Barbosa:2004kp,Guzman:2007}. The main purpose of this letter is to analyze the effects of more general phase space deformations in cosmology. In \cite{darabi} the authors study effects of the more general deformations of the minisuperspace of dilation cosmology, they comment that in the late time behavior of the model is similar to that of a de Sitter universe.

We will work with an FRW universe and  a scalar field. This model has been used to explain several aspects of our universe, like inflation, dark energy and dark matter. Our approach to deformed space cosmology is through its introduction in a phase space constructed in the minisuperspace variables, and is achieved by modifying the symplectic structure (Poisson's algebra of the minisuperspace) in the same manner as in \cite{Barbosa:2004kp,Pimentel:2004jv,Guzman:2007}. It will be showed that  in the absence of a cosmological constant, the behavior of the scale factor can account for a late time acceleration.

We will start in Section II, by introducing the commutative model. In Section III, the noncommutative model is presented, as well as the dynamics of the cosmological model. We will show that with our approach late time acceleration can be accounted for  and a relationship between the cosmological constant and the deformation parameters is conjectured. Finally, the last section is devoted for conclusions and outlook.

\section{The Commutative Model}
As already suggested, cosmology presents an attractive arena for noncommutative models, both in the quantum as well as classical level. One of the features of noncommutative field theories is UV/IR mixing, this effectively mixes short scales with long scales, from this fact one may expect that even if  noncommutativity is present at a really small scale, by this UV/IR mixing, the effects might be present at an older time of the universe.

We start with  a homogeneous and isotropic universe endowed with  a flat Friedmann-Robertson-Walker  metric 
\begin{equation}
ds^2 = -N^2(t) dt^2+ a^{2}(t)\left[ dr^2 +r^2d\Omega\right],
\end{equation}
where as usual $a(t)$ is the scale factor of the universe and $N(t)$ is the lapse function. The action we will be working with is the Einstein-Hilbert action and a scalar field $\phi$ as the matter content for the model. 
In units $8\pi G=1$, the action takes the form
\begin{equation}
S=\int dt \left\{  -\frac{3a\dot{a}^{2}}{N}+a^{3}\left( \frac{\dot{\phi}^{2}}{2N}+N V(\phi) \right) \right\},   \label{susy20}%
\end{equation}
here $V(\phi)$ is the scalar potential. From now on we will restrict to the case of   constant potential ($V(\phi)=-\Lambda$).

 For the purposes of introducing the deformation to the minisuperspace variables an appropriate redefinition needs to be made, we make the following change of variables
\begin{equation} 
 x = m^{-1} a^{3/2} \sinh ( m \phi ),~ y = m^{-1} a^{3/2} \cosh (m \phi). 
\end{equation}
 where $m^{-1}=2\sqrt{2/3}$. In these new variables we calculate the Hamiltonian 
\begin{equation}
 H_c =N \left( \frac{1}{2} P_x^2 + \frac{\omega^2}{2} x^2\right) -N \left( \frac{1}{2} P_y^2 + \frac{\omega^2}{2} y^2 \right),  \label{canonico1} 
\end{equation}
where $\omega^2=-\frac{3}{4}\Lambda$. This is the canonical Hamiltonian which is a first-class constraint  as is usual in general relativity. Since we do not have second class constraints in the model we will continue to work with the usual   Poisson brackets and the relations of commutation between the phase space variables 
\begin{equation}
\{ x_i, x_j \} = 0, \qquad \{ P_{x_i} , P_{y_j} \} = 0,\qquad\{x_i, P_{x_j} \}  = \delta_{ij}. \label{classicPoisson}
\end{equation}
At this point, we have a minisuperspace spawned by the new variables $(x,y)$. To find the dynamics of this model, we need to solve the equations of motions, these are derived as usual by using Hamiltons equations.  For the particular model the equations are
\begin{eqnarray}
\dot x&=&-P_x,\qquad \dot y=P_y,\\
\dot P_x&=&\omega^2 x,\qquad \dot P_y=-\omega^2 y,\nonumber
\end{eqnarray}
these equations can easily be integrated 
\begin{equation}
x(t)=X_0\cos{(\vert\omega\vert t+\delta_x)},\quad y(t)=Y_0\cos{(\vert\omega\vert t+\delta_y)}.\\
\end{equation}
In order to satisfy the Hamiltonian constraint we introduce the solutions to the Hamiltonian, this gives a relationship between the integration constants, it easy to verify that $X_0=\pm Y_0$. 
\section{Deformed Space Model and $\Lambda$}
The original ideas of deformed phase space, or more precisely deformed minisuperspace, where done in conection with noncommutative cosmology \cite{GarciaCompean:2001wy}. As already mentioned, in order to avoid the complications of a noncommutative theory of gravity, they introduce a deformation to the minisuperspace in order to incorporate noncommutativity. Usually the deformation is introduced by the Moyal brackets, which is based in the Moyal product. An alternative to introduce full phase space deformations can be accomplished by deformation quantization \cite{hugo}. In quantum cosmology, the models are constructed in the minisuperspace, then we can argue that studying cosmology in deformed phase space could be related to studying quantum effects to cosmological models \cite{darabi}. A different constructio is based on symplectic manifolds \cite{libro}. In this approach  the Hamiltonian has the same functional form but is valued  on variables that satisfy a modified Poisson algebra \cite{Barbosa:2004kp,libro, darabi}.
Once the deformation has been done one arrives to a modified Poisson algebra
\begin{equation}
\{x_i,x_j\}_{\alpha}=\theta_{ij}, \quad \{x_i,p_j\}_{\alpha}=\delta_{ij}+\sigma_{ij},\quad \{p_i,p_j\}_{\alpha}=\beta_{ij}.
\end{equation}
Making the following transformation on the classical phase space variables $\{x,y,p_x,p_y\}$
\begin{eqnarray}
\hat{x}=x+\frac{\theta}{2}p_{y}, \qquad \hat{y}=y-\frac{\theta}{2}p_{x},\nonumber \\
\hat{p}_{x}=p_{x}-\frac{\beta}{2}y, \qquad \hat{p}_{y}=p_{y}+\frac{\beta}{2}x, \label{nctrans}
\end{eqnarray}
now the algebra reads
\begin{equation}
\{\hat{y},\hat{x}\}=\theta,\quad \{\hat{x},\hat{p}_{x}\}=\{\hat{y},\hat{p}_{y}\}=1+\sigma,\quad \{\hat{p}_y,\hat{p}_x\}=\beta,\label{dpa}
\end{equation}
where $\sigma=\theta\beta/4$.
Now that we have constructed the modified phase space, we apply the transformation to the Hamiltonian (\ref{canonico1}), where after defining
\begin{equation}
\omega_{1}^2=\frac{\beta-\omega^2\theta}{1-\omega^2\theta^2/4}, \quad\omega_{2}^2=\frac{\omega^2-\beta^2/4}{1-\omega^2\theta^2/4},\label{omegas}
\end{equation}
we get
\begin{equation}
\widehat{H}= \frac{1}{2}\left \{ \hat p_{x}^2-\hat p_{y}^2+\omega_{1}^2(\hat x\hat p_{y}+\hat y\hat p_{x}) + \omega_{2}^2(\hat x^2-\hat y^2) \right\} . \label{HNC}
\end{equation}
We can construct a bidimensional vector potential $\hat{A}_{x}=-\frac{\omega_1^2}{2}\hat{y}$, $\hat{A}_{y}=\frac{\omega_1^2}{2}\hat{x}$ from were a magnetic field $B=\omega_1^2$ is calculated, this result allow us to write the effects of the noncommutative deformation as minimal coupling on the Hamiltonian, $\widehat{H}=[(p_{x}-\hat{A}_{x})^2+\omega_3^{ 2}\hat{x}^2]-[(p_{y}-\hat{A}_{y})^2+\omega_3^{2}\hat{y}^2]$, this expression can be rewritten in terms of the magnetic B-field as

\begin{eqnarray}
\widehat{H}&=&[\:\hat{p}_{x}^2+(\omega^{2}_{3}-B^2/4)\hat{x}^2]-[\:\hat{p}_{y}^2+(\omega^{2}_{3}-B^2/4)\hat{y}^2]\nonumber\\
&+&B(\hat{y}\hat{p}_{x}+\hat{x}\hat{p}_{y}),
\end{eqnarray}
where $\omega^{2}_{3}=\omega_{1}^2+\omega_{2}^2$. This is a typical result in 2-dimensional noncommutativity, where the effects of the noncommutative deformation can be encoded in a perpendicular constant magnetic field. 

To obtain the dymanics for our model, we can derive the equations of motion from the Hamiltonian (\ref{HNC})
\begin{eqnarray}\nonumber
\dot{\hat x}=\{\hat{x},\widehat{H}\}=\frac{1}{2}[2\hat p_{x}+\omega_{1}^2\hat{y}],\\ \nonumber
\dot{\hat y}=\{\hat{y},\widehat{H}\}=\frac{1}{2}[-2\hat p_{y}+\omega_{1}^2\hat{x}],  \nonumber \\
{\dot{\hat p}_{x}}=\{\hat p_{x},\widehat{H}\}=\frac{1}{2}[-\omega_{1}^2\hat p_{y}-2\omega_{2}^2\hat{x}], \\ \nonumber
{\dot{\hat p}_{y}}=\{\hat p_{y},\widehat{H}\}=\frac{1}{2}[-\omega_{1}^2\hat p_{x}+2\omega_{2}^2\hat{y}],\nonumber \label{nceqm}
\end{eqnarray}
yielding the following equations of motion
\begin{eqnarray}
\ddot{x}-{\omega}^{2}_1\dot{y}+\left(\omega^2_2+\frac{1}{4}\omega^4_1\right)x=0,\nonumber \\
\ddot{y}-{\omega}^{2}_1\dot{x}+\left(\omega^2_2+\frac{1}{4}\omega^4_1\right)y=0.\label{15}
\end{eqnarray}
Defining new variables $\eta$ and $\zeta$ as $\eta=\hat x+\hat y$, $\zeta=\hat y-\hat x$, we can decouple and solve equations (\ref{15}) to get  the solutions in terms of the noncomutative variables $\hat{x}(t)$ and $\hat{y}(t)$

\begin{equation}\nonumber
\hat{x}(t)=A\:e^{\frac{\omega_{1}^2}{2}t} \cosh\left(\omega^{\prime}t+\delta_1\right)-B\:e^{-\frac{\omega_{1}^2}{2}t} \cosh\left(\omega^{\prime}t+\delta_2\right),\label{xnc}
\end{equation}
\begin{equation}
\hat{y}(t)=A\:e^{\frac{\omega_{1}^2}{2}t} \cosh\left(\omega^{\prime }t +\delta_1\right)+ B\:e^{-\frac{\omega_{1}^2}{2}t} \cosh\left(\omega^{\prime }t+\delta_2 \right),\label{ync}
\end{equation}
where $\omega^{\prime 2}=-\omega_{2}^2.$

Up to this point we have obtained the equations of motion using the deformed Poisson algebra (\ref{dpa}). From the solutions for (\ref{ync}) we can write  the volume of the universe original variables,
\begin{eqnarray}
a^3(t)&=&V\left[2\cosh^2\left(\omega^{\prime}t+\frac{\delta_1+\delta_2}{2}\right)\right.\nonumber\\
& &\left.-1+\cosh\left(\delta_1-\delta_2\right)\right]\label{eq17}\nonumber\\
\end{eqnarray}
 where $\delta_1$ and $\delta_2$ are related to the constants $A,B$. From Figure 1 we can notice that for large values of $t$  our cosmological model behaves like a de Sitter cosmology.  
 Comparing the de Sitter cosmology with the deformed phase space model in the late time limit enables us to get the following relationships between the de Sitter cosmological constant $\Lambda$ and the noncommutative parameters
 \begin{equation}
\Lambda_{eff} =\frac{1}{3}\left(\frac{\beta^2+3\Lambda}{1+\frac{3}{16}\Lambda\theta^2}\right ),
\end{equation}
in the particular case where $\Lambda=0$, the noncommutative parameter plays the role of the cosmological constant and is given by

\begin{equation}
\Lambda_{eff}=\frac{\beta^2}{3}.
\end{equation}

\begin{figure}
\begin{center}
\includegraphics[width=8cm]{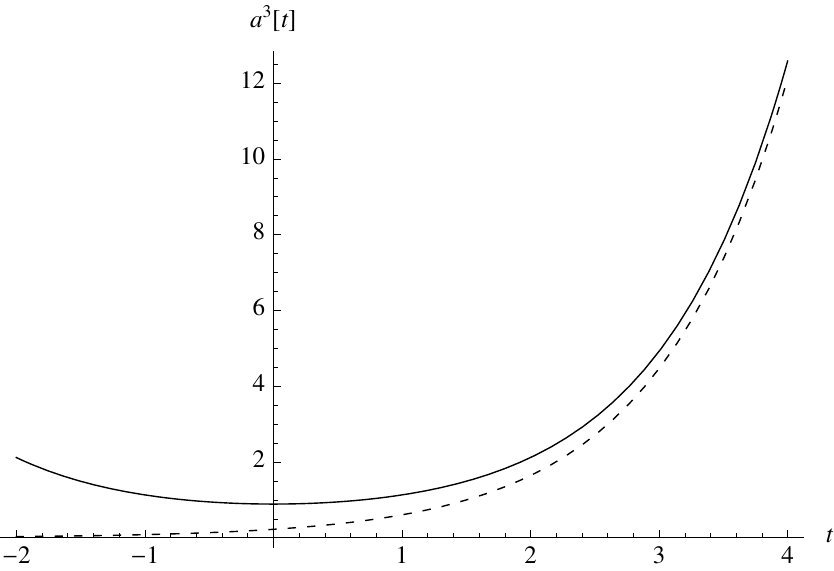}
\end{center}
\caption{\label{fig:de} 
Dynamics of the phase space deformed model for the values $X_0=Y_0=1,\delta_2=\delta_1=0$, $\omega=0$ and $\beta=1$. The solid line corresponds to the volume of the universe, calculated with the noncommutative model. The dotted line corresponds to the volume of the de Sitter spacetime. For large values of $t$ the behavior is the same.}
\end{figure}

\section*{Discussion and Outlook}

 In this letter we have constructed a deformed phase space model of scalar field cosmology. The deformation is introduced by modifying the symplectic structure of the minisuperspace variables, in order to have a deformed Poisson algebra among the coordinates and  momenta. This construction is consistent with the assumption taken in noncommutative quantum cosmology \cite{GarciaCompean:2001wy,Barbosa:2004kp,Pimentel:2004jv,Guzman:2007}, and enable us to study the effects of phase space deformations in scalar field cosmology. 

The deformed phase space model is obtained making the transformation (\ref{nctrans}) on the canonical Hamiltonian (\ref{canonico1}) which allow us to work out a Hamiltonian that depends on the deformed variables $\hat{x_i}$ and $\hat{p_i}$. To obtain the noncommutative equations of motion for the model, an in order to find solutions, a convenient change of variables was made. Finally, with the solutions, we where able to return to the original variables, and find that the volume of the universe is given by equation (\ref{eq17}).

We found interesting  effects on the evolution of the scale factor as a consequence of the deformation to the phase space.  First, in the case  when we turn off the parameter $\theta$ in (\ref{omegas}), the noncommutative parameter $\beta$ can be interpreted as a magnetic field that is constructed from a 2 dimensional vector potential. The effects of this B-field can be introduced in to the Hamiltonian  through minimal coupling on the canonical momenta. 

Finally we found that with our model the volume of the universe behaves like a de Sitter cosmology for large values of $t$ even when $\Lambda=0$. This allows us to get a relation between the cosmological constant and the deformed parameter for the momenta by comparing the late time evolution of the volume of the noncommutative model with the volume of a de Sitter universe. Evidence of a possible relationship between the late time acceleration of the universe and the noncommutative parameters has been accumulating \cite{darabi,lambda,quiros}, our results also point in this direction, based on this observation in  a model that gives some insight of the origin of $\Lambda$ is presented in \cite{lambda2}.

\section*{Acknowledgments}
This work is  supported by DAIP grant 18/10 and  CONACYT grants 62253, 84798, 135023. S.P.P is supported by CONACyT PhD. grant.


\begin{thebibliography}{99}
 \bibitem{nekra} N.~Seiberg and E.~Witten,
  JHEP {\bf 9909}, 032 (1999); A. Connes, M. R. Douglas, and A. Schwarz, {JHEP} {\bf %
9802:003} (1998); M.~R.~Douglas and N.~A.~Nekrasov,
  Rev.\ Mod.\ Phys.\  {\bf 73}, 977 (2001).
 \bibitem{GarciaCompean:2003pm}
  H.~Garcia-Compean, O.~Obregon, C.~Ramirez and M.~Sabido,
  Phys.\ Rev.\  D {\bf 68}, 044015 (2003); H.~Garcia-Compean, O.~Obregon, C.~Ramirez and M.~Sabido,
  Phys.\ Rev.\  D {\bf 68}, 045010 (2003);
  P.~Aschieri, M.~Dimitrijevic, F.~Meyer and J.~Wess,
  Class.\ Quant.\ Grav.\  {\bf 23}, 1883 (2006);
  X.~Calmet and A.~Kobakhidze,
  Phys.\ Rev.\  D {\bf 72}, 045010 (2005);
  L.~Alvarez-Gaume, F.~Meyer and M.~A.~Vazquez-Mozo,
  Nucl.\ Phys.\  B {\bf 753}, 92 (2006); S.~Estrada-Jimenez, H.~Garcia-Compean, O.~Obregon and C.~Ramirez,
  Phys. Rev. D {\bf 78},124008 (2008).
  \bibitem{GarciaCompean:2001wy}
  H.~Garcia-Compean, O.~Obregon and C.~Ramirez,  Phys.\ Rev.\ Lett.\  {\bf 88}, 161301 (2002).
\bibitem{Barbosa:2004kp}
  G.~D.~Barbosa and N.~Pinto-Neto,  Phys.\ Rev.\  D {\bf 70}, 103512 (2004).
\bibitem{Pimentel:2004jv}
  L.~O.~Pimentel and C.~Mora,
  Gen.\ Rel.\ Grav.\  {\bf 37} (2005) 817;  L.~O.~Pimentel and O.~Obregon,
  Gen.\ Rel.\ Grav.\  {\bf 38}, 553 (2006); M.~Aguero, J.~A.~Aguilar S., C.~Ortiz, M.~Sabido and J.~Socorro,
  Int.\ J.\ Theor.\ Phys.\  {\bf 46}, 2928 (2007);
 W.~Guzman, C.~Ortiz, M.~Sabido, J.~Socorro and M.~A.~Aguero,
  Int.\ J.\ Mod.\ Phys.\  D {\bf 16}, 1625 (2007);B.~Vakili, N.~Khosravi and H.~R.~Sepangi,
  Class.\ Quant.\ Grav.\  {\bf 24} (2007) 931
     \bibitem{Guzman:2007}
  W.~Guzman, M.~Sabido and J.~Socorro, 
  Phys.\ Rev. \ D {\bf 76}, 087302 (2007).
  %
\bibitem{hugo} R.~Cordero, H.~Garcia-Compean and F.~J.~Turrubiates,
  Phys.\ Rev.\ D {\bf 83}, 125030 (2011)
  [arXiv:1102.4379 [hep-th]].
\bibitem{darabi}
B.~Vakili, P.~Pedram, S.~Jalalzadeh,
  Phys.\ Lett.\  {\bf B687}, 119-123 (2010).
  
 \bibitem{libro} W.~Guzman, M.~Sabido, J.~Socorro,
  Phys.\ Lett.\  {\bf B697}, 271-274 (2011);V.~Guillemin and S.~Sternberg,
{\it  Cambridge, UK: Univ. Pr. (1990) 468 p}.

 \bibitem{lambda} O.~Obregon, M.~Sabido, E.~Mena,
  Mod.\ Phys.\ Lett.\  {\bf A24}, 1907-1914 (2009).
  \bibitem{quiros} O.~Obregon, I.~Quiros,
  Phys.\ Rev.\  {\bf D84}, 044005 (2011).
\bibitem{lambda2} S. P\'erez-Pay\'an, E. Mena and M. Sabido, to appear.
  \end{thebibliography}
  \end{document}